# Electrodynamics of the Josephson-Coupled Parallel Plate Resonator

Vladimir V. Talanov[*]

*Neocera, LLC, Beltsville, MD 20705*

Eigen oscillations in a superconducting parallel plate resonator with the Josephson-coupled plates are investigated. While the insulator thickness $S$ changes from tens of microns down to the decay lengthscale of the superconducting wavefunction into a dielectric, $\xi_d \sim 1$ nm, both the resonant frequency and $Q$-factor vary non-monotonically by up to three orders in magnitude. A crossover between the Swihart waves and Josephson plasmons causes a global minimum in the resonant frequency and a local maximum in the $Q$-factor at $S \gg \xi_d$.

Electrodynamic response of a superconductor provides much insight into its intrinsic properties, such as carrier density, pairing state symmetry, or quasiparticle excitation spectrum [1]. Superconducting waveguides and resonators have been extensively explored, both theoretically and experimentally, in conjunction with the microwave characterization of various superconductors, especially high-$T_c$ thin films [2], [3]. Conceivably, the most common geometry has been a superconductor-insulator-superconductor (SIS) sandwich [4]-[15], shown in Fig. 1. Its eigen oscillations are associated with the Swihart wave [4], a TM mode slowed down due to disparity between the spatial extend of the electric and magnetic fields, with the phase velocity

$$v_S = c\big/\sqrt{\varepsilon_d\left(1+2\lambda_{\mathit{eff}}/S\right)} \qquad (1)$$

Here $c$ is the speed of light in vacuum, $\varepsilon_d$ is the insulator relative dielectric constant, $\lambda_{\mathit{eff}} = \lambda \coth(d/\lambda)$ is the effective penetration depth into the superconducting film of thickness $d$ with $\lambda$ the intrinsic penetration depth, and $S$ is the insulator thickness. Taber [6] introduced a parallel plate resonator (PPR), formed by 10- to 100-μm-thick dielectric spacer, sandwiched between two superconducting films on dielectric substrates of 1×1 cm in size (see Fig. 1). PPR has a resonant frequency and *Q*-factor of about 10 GHz and few thousands, respectively, and has been very fruitful for measuring the superconductor impedance as a function of temperature [7], magnetic field [8], film thickness [9]; optimizing the film growth conditions [10]; investigating the superconductor/ferromagnet heterostructures [11] and non-linear properties of superconductors [15]; and for measuring the London penetration depth [12].

---

[*] E-mail: talanov@neocera.com

Noticeably, at a nanometer thin dielectric PPR resembles a SIS tunnel junction. The junction eigen oscillations are associated with Josephson plasmons [16], first observed by Dahm, *et al*. [17], where the reactive energy oscillates between the barrier electric energy and the inductive energy of the superconductive condensate, while $v_s$ represents a minimum phase velocity of linear plasma waves propagating in the system. The Josephson plasma frequency is

$$\omega_J = v_S/\lambda_J = \sqrt{2\pi J_c/\Phi_0 C} \tag{2}$$

where $\lambda_J = \sqrt{\Phi_0/2\pi\mu_0(S+2\lambda_{eff})J_c}$ is the Josephson penetration depth, $J_c$ is the Josephson critical current density, $\Phi_0 = \pi\hbar/e$ is the magnetic flux quantum with $\hbar$ the reduced Planck constant and *e* the electron charge, $\mu_0$ is the permeability of free space, and $C = \varepsilon_0\varepsilon_d/S$ is the junction capacitance per unit area with $\varepsilon_0$ the permittivity of free space.

To date, there have been numerous publications devoted, on one hand, to Swihart waves in SIS waveguides and resonators without tunneling [4]-[15], and, on the other hand, to Josephson plasmons in SIS tunnel junctions [16]-[18] and their stacks [19]-[22]. At the same time, a single theory addressing the evolution of eigen oscillations in a Josephson-coupled SIS sandwich versus its principal geometric parameter, insulating barrier thickness *S*, has not been published. Therefore, the purpose of this paper is to calculate a resonant frequency and *Q*-factor of *tunnel* PPR (TPPR) versus *S* continuously varying from ten micron down to the decay lengthscale of the superconducting wavefunction into a dielectric spacer, $\xi_d \sim 1$ nm. We also study in detail a dimensional crossover between the two types of oscillations, the Swihart waves and Josephson plasmons, governing TPPR behavior at thicker and thinner dielectric spacings, respectively.

Let us recollect a theory of conventional PPR [6], [12]. Ignoring the open-end correction, the Swihart mode resonant frequency is

$$\omega_S = n\pi v_S/L \tag{3}$$

where $n=1, 2, 3 \ldots$ is the mode index, and $L$ is the resonator length. For dielectric spacings thinner than 10-20 μm the $Q$-factor is controlled by the ohmic losses:

$$Q_S = Q_{S\Omega} = \frac{\omega_S \mu_0 S}{2R_{eff}} + \frac{\omega_S \mu_0 \lambda_{eff}}{R_{eff}} \tag{4}$$

Here,

$$R_{eff} = \left(\frac{\omega}{\omega^*}\right)^2 \left( R_s^* \left( \coth\frac{d}{\lambda} + \frac{d/\lambda}{\sinh^2(d/\lambda)} \right) + \frac{(\omega^* \mu_0 \lambda)^2}{\sinh^2(d/\lambda)} \sqrt{\frac{\varepsilon_0 \varepsilon_{sub}}{\mu_0 \mu_{sub}}} \right) \tag{5}$$

is the frequency dependent effective surface resistance of the thin superconducting film [23] on a substrate with relative permittivity $\varepsilon_{sub}$ and permeability $\mu_{sub}$, and $R_s^*$ is the intrinsic surface resistance at fixed frequency $\omega^*$. The second addend in the right hand side of (4) is due to the resonator energy stored in both the geometric and kinetic inductances of the superconducting plates [9]; because of $\omega \mu_0 \lambda_{eff} \gg R_{eff}$ PPR retains $Q \gg 1$ even at very thin dielectrics of $S \ll \lambda_{eff}$.

Let us now consider a TPPR. The eigen waves in SIS waveguide obey the perturbed sine-Gordon equation for the phase difference $\Theta$ between the superconducting quantum mechanical wavefunctions [24]:

$$\frac{\partial^2 \Theta}{\partial t^2} - v_S^2 \frac{\partial^2 \Theta}{\partial x^2} + \omega_J^2 \sin\Theta = \alpha \frac{\partial^3 \Theta}{\partial x^2 \partial t} - \beta \frac{\partial \Theta}{\partial t} \tag{6}$$

where $\omega_J$ is given by (2) or (A5), $\alpha = 1/R_{sc}C$ is the attenuation coefficient due to the ohmical losses in the superconducting plates, $R_{sc} = \omega^2 \mu_0^2 (S + 2\lambda_{eff})^2 / 2R_{eff}$ is the superconductor equivalent resistance [25], $\beta = 1/R_{qp}C$ is the attenuation coefficient due to the quasiparticle tunneling, with $R_{qp}$ the quasiparticle tunneling resistance per unit area. Although, high quality SIS junctions exhibit $R_{qp} \approx R_n \exp(\Delta/k_B T)$, we assume that $R_{qp} \simeq R_n$ which is typically observed in practice [26], with $R_n$ the normal state tunneling resistance given by (A2), and $k_B$ the Boltzmann constant. We restrict ourselves to small oscillations of $\Theta$ around zero. Substitution of $\sin \Theta \simeq \Theta$ and time-space dependence $\exp(i\omega t - i\gamma x)$ into (6) and yields for the propagation constant $\gamma = \gamma_1 - i\gamma_2$:

$$\gamma = \frac{\sqrt{\omega^2 - \omega_J^2}}{v_S} - i\left(\frac{\alpha\omega\sqrt{\omega^2 - \omega_J^2}}{2v_S^3} + \frac{\beta\omega}{2v_S\sqrt{\omega^2 - \omega_J^2}}\right) \tag{7}$$

where only the leading terms due to $\gamma_2 \ll \gamma_1$ are retained. To find TPPR resonant frequency and $Q$-factor, we employ a resonant condition [12]

$$\Gamma^2 \exp(-2i\gamma L) = \exp(-2i\pi n) \tag{8}$$

where $\Gamma$ is the reflection coefficient from the open end, which can be set to $-1$ because of $S \ll L$, and $n=1, 2, 3 \dots$ . Substituting (7) and complex angular frequency $\omega = \omega_1 + i\omega_2$ into (8), retaining only the leading terms due to $\gamma_2 \ll \gamma_1$, $\omega_2 \ll \omega_1$, $\alpha\gamma^2 \ll \omega_1$ and $\beta \ll \omega_1$, and separating the real and imaginary parts yields for the resonant frequency $\omega_{res} = \omega_1$ and quality factor $Q = \omega_1/2\omega_2$:

$$\omega_{res} = \sqrt{\omega_S^2 + \omega_J^2} \tag{9}$$

$$\frac{1}{Q} = \frac{1}{Q_\Omega} + \frac{1}{Q_{qp}} = \frac{2\omega_S^2 R_{eff}}{\omega_{res}^3 \mu_0 (S + 2\lambda_{eff})} + \frac{1}{\omega_{res} R_{qp} C} \tag{10}$$

The first and second addends in the right hand side of (10) represent the ohmical and quasiparticle tunneling losses, respectively (note, that $Q_{qp}^2$ is identical with the Stewart-McCumber parameter [27]). Without the tunneling, (9) and (10) reduce to (3) and (4), respectively.

Figure 2 shows TPPR resonant frequency and $Q$-factor for two lowest modes ($n=1, 2$) versus normalized dielectric spacer thickness $S/\xi_d$. The following representative parameters are used: $L$=10 mm, $d$=200 nm, $\lambda$=150 nm, $R_s^* = 100\ \mu\Omega$ at $\omega^* = 10$ GHz, the normal state conductivity $\sigma_n$=0.01 $(\mu\Omega\cdot\text{cm})^{-1}$, $\varepsilon_d$=4, $\xi_d$=1 nm (*e.g.*, see [28]), $\varepsilon_{sub}$=10, and $\mu_{sub}$=1. In the limit $S \gg \lambda_{eff}$ the resonant frequency approaches $n\pi c/\sqrt{\varepsilon_d} L$, that is a frequency of perfectly conducting PPR. Decreasing $S$ reduces $\omega_{res}$ due to growing disparity between the spatial extend of electric and magnetic fields. Between $S/\xi_d \sim 10$ and ~20 $\omega_{res}$ reaches a global minimum (see Fig. 3a) since the tunneling becomes strong enough for the Josephson plasmons to conquer the Swihart waves. Finally, at $S \ll S_0$ the resonant frequencies of all modes converge to $\omega_J$ exponentially dependent on $S$. The dielectric spacer thickness $S_0$ associated with the Swihart-Josephson crossover can be found from $\partial \omega_{res}/\partial S = 0$:

$$S_0 = \frac{\xi_d}{\sqrt{2}} \ln \frac{3eL^2}{4\pi^3 n^2 \Phi_0 \sigma_n \lambda \xi_d^2} \tag{11}$$

Using the above quantities we find $S_0 \approx 15\xi_d$. Substituting (11) into (A5) reveals that TPPR transforms from a short junction ($L < \lambda_J$) at $S > S_0$ into a long junction ($L > \lambda_J$) at $S < S_0$.

The $Q$-factor dependence has four distinct regimes (see Fig. 2b). At thick dielectrics of $S \gg \lambda_{eff}$ we observe $Q \to \omega_S \mu_0 S/2R_{eff}$. Reducing $S$ decreases the resonant frequency, which, in turn, reduces the ohmical losses due to $R_{eff} \propto \omega_{res}^2$. At $S = 2\lambda_{eff}$ the magnetic energy becomes equally stored in the spacer geometric inductance, and geometric and kinetic inductances of the superconductive plates. For $S_0 < S \ll 2\lambda_{eff}$ spacings, the magnetic energy is stored in the geometric and kinetic inductances of the plates. Interplay of these phenomena yield a broad minimum in $Q$ about $S \sim \lambda_{eff}$ ($S/\xi_d \sim 100\text{-}200$) followed by its rise. Just below $S_0$ $Q$ continues to rise even more rapidly due to an extra inductive energy brought in by the tunneling Cooper pairs (see Fig. 3b), while the ohmical losses still dominate. Further reduction of $S$ brings the quasiparticle tunneling into play, creating a maximum in $Q$ around $S/\xi_d \sim 12$ where the dissipation is due equally to the ohmical and quasiparticle tunneling losses. Finally, the quasiparticle tunneling makes $Q \simeq Q_{qp}$ to reduce abruptly. Because at temperatures not too close to $T_c$ the Cooper pair density is much higher than the quasiparticle density, there is a few-$\xi_d$-wide range of dielectric spacings, just below $S_0$, where the Josephson plasmons are unaffected by the quasiparticle tunneling.

Let us discuss TPPR behavior as $S/\xi_d$ approaches unity. Both the local electrodynamics condition $\lambda_J \gg S + 2\lambda_{eff}$ and Wentzel-Kramers-Brillouin approximation, casual to (6) and (A1), respectively, remain valid down to $S \geq \xi_d$. For plates made of conventional superconductor with the gap frequency $\omega_\Delta \simeq 2\Delta/\hbar$ much lower than the bulk plasma frequency $\omega_p = c/\lambda$, the frequency of eigen oscillations may not exceed $\omega_\Delta$, for instance 1 THz at $S/\xi_d \sim 3$ (see solid lines in Fig. 2). Below such spacing the superconductivity is destroyed and the oscillations are overdumped. A qualitatively different scenario takes place in the case of strongly anisotropic superconductor ($c \parallel z$, see Fig. 1), in which $c$-axis plasma frequency $\Omega_J$ (*e.g.*, again 1 THz) lays well below $\omega_\Delta$. Once $\omega_{res} = \omega_J$ exceeded $\Omega_J$ the oscillations are no longer confined within the junction because of $c$-axis plasma excitations formed in the plate bulk [21], although no substantial dumping is introduced. For the sake of clarity, in Fig. 2 we extend the dependences given by (9) and (10) down to $S/\xi_d = 1$ by dashed lines.

Although, fabricating TPPR with variable barrier thickness may be rather challenging (*cf.*, [12]), we hypothesize that a Swihart-Josephson crossover can be experimentally observed by pressing two superconducting films face-to-face and measuring the resonator frequency and *Q*-factor as the functions of pressure. A native oxide and perhaps Cu outgrowths (*e.g.*, see [10]) present on the film surface may create between the plates a plurality of weak-link Josephson junctions, resembling the distributed junction behavior. More elaborate experiment would be investigating multiple TPPRs of the same length with barrier thickness ranging from $S \gg S_0$ down to $S \geq \xi_d$, which may be technologically feasible even for 1-cm-long junctions due to relatively thick crossover spacing (see (11)). The microwave radiation can be coupled into such a structure by antenna probes devised in [12].

To conclude, we derived the expressions for a resonant frequency and *Q*-factor of superconducting tunnel parallel plate resonator, which are applicable over the entire range of physically important dielectric spacings, *e.g.*, from tens of microns down to the decay lengthscale of the superconducting wavefunction into a dielectric barrier, $\xi_d$. A global minimum in the resonant frequency, due to a crossover between the Swihart wave and Josephson plasma oscillations, occurs for 1-cm-long resonator at the spacer thickness $S_0 \sim 15\xi_d$. An unanticipated maximum in *Q*-factor at barrier thickness few $\xi_d$ below $S_0$ is due to interplay between the ohmical and quasiparticle tunneling losses.

This work has been supported by NSF-SBIR IIP-0924610. Author is thankful to Prof. S. Anlage and Dr. V. Borzhenets for critical comments on the manuscript.

**Appendix**

To derive the barrier thickness dependence for a normal state tunneling resistance of SIS tunnel junction, consider a metal-insulator-metal tunnel junction. Assuming a rectangular potential barrier, same electrodes, and same electron mass *m* in both the barrier and electrodes, the tunnel current density at small bias voltage $V \simeq 0$ is [29]:

$$J = \frac{3e^2 \kappa V}{8\pi^2 \hbar S} \exp(-2\kappa S) \tag{A1}$$

where *e* is the electron charge, $\kappa = \sqrt{2m\varphi_0}/\hbar$ is the electron wavefunction decay rate into the barrier, with *m* the electron mass and $\varphi_0$ the barrier height. Since the Cooper pair of mass 2*m* and single electron of mass *m* have the same tunneling probability, one may conclude that quantity $1/\sqrt{2}\kappa$ resembles a superconducting wavefunction decay lengthscale into a dielectric barrier, that is $\xi_d = 1/\sqrt{2}\kappa = \hbar/\sqrt{4m\varphi_0}$ (*cf.*, the Ginzburg-Landau coherence length $\xi_c = \hbar/\sqrt{4m|a(T)|}$ with $a(T)$ the phenomenological coefficient in the free energy density expansion). Substitution of $\kappa = 1/\sqrt{2}\xi_d$ into (A1) yields for the tunneling resistance per unit area $R_n = V/J$:

$$R_n = \frac{8\sqrt{2}\pi^2 \hbar \xi_d S}{3e^2} \exp\left(\frac{\sqrt{2}S}{\xi_d}\right) \tag{A2}$$

To obtain the barrier thickness dependence for the Josephson plasma frequency $\omega_J$, we invoke the Ambegaokar-Baratoff relation for the critical current density of SIS junction [30]

$$J_c = \frac{\pi \Delta(T)}{2eR_n} \tanh \frac{\Delta(T)}{2k_B T} \tag{A3}$$

and the Mattis-Bardin expression for the imaginary part $\sigma_{sc} = 1/\mu_0 \omega \lambda^2$ of the superconductor complex conductivity [31]

$$\sigma_{sc} = \sigma_n \frac{\pi \Delta(T)}{\hbar \omega} \tanh \frac{\Delta(T)}{2k_B T} \tag{A4}$$

where $R_n$ is given by (A2), and $\sigma_n$ is the normal state conductivity. Since (A4) is applicable only at frequencies below the energy gap, we suppose that $\sigma_n$ is frequency independent. From (2), (A2), (A3) and (A4) we obtain:

$$\omega_J = \left(\frac{3e^2}{8\sqrt{2}\pi^2\hbar\xi_d\sigma_n}\right)^{1/2}\frac{\omega_p}{\sqrt{\varepsilon_d}}\exp\left(-\frac{S}{\sqrt{2}\xi_d}\right) \tag{A5}$$

where $\omega_p = c/\lambda$ is the superconductor bulk plasma frequency. Since the prefactor in (A5) is less than unity (e.g., $\left(3e/8\sqrt{2}\pi\Phi_0\sigma_n\xi_d\right)^{1/2} \sim 0.1$ for the representative values of $\sigma_n$=0.01 (μΩ·cm)$^{-1}$ and $\xi_d$=1 nm), we notice that for $\exp\left(-S/\sqrt{2}\xi_d\right) \ll 1$, as expected, $\omega_J \ll \omega_p$.

**Figures**

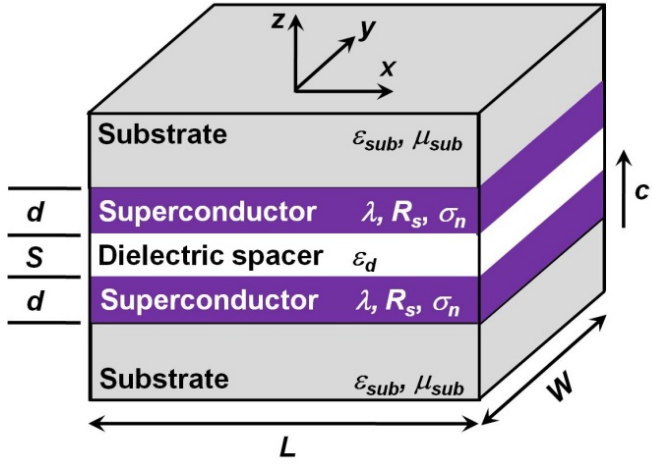

**Figure 1.** SIS sandwich with $S \ll L \sim W$. A Swihart wave with the field components $E_z$, $H_y$, and $E_x \ll E_z$ propagates along $x$-axis.

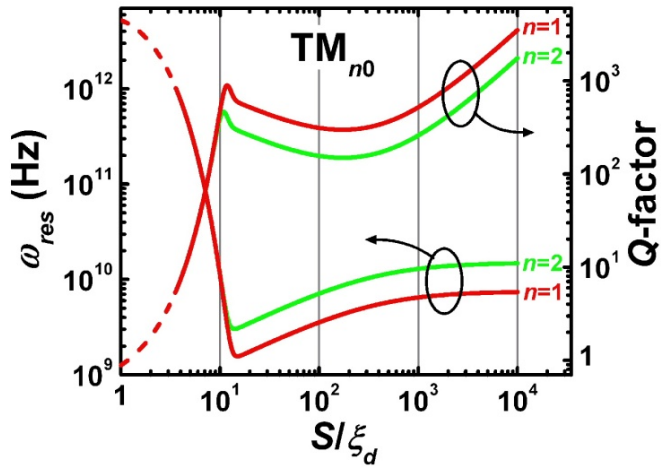

**Figure 2.** Log-log plot of the resonant frequency $\omega_{res}$ and $Q$-factor given by (9) and (10), respectively, versus normalized dielectric spacer thickness $S/\xi_d$ for $TM_{10}$ and $TM_{20}$ modes.

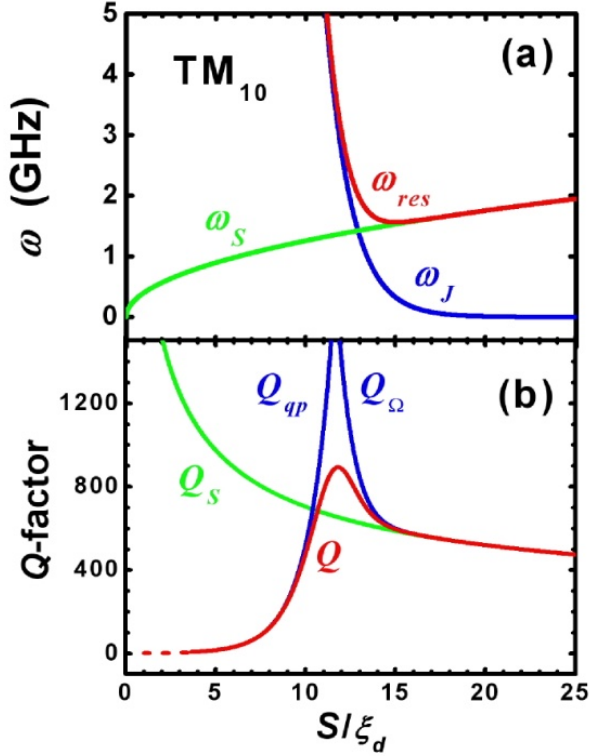

**Figure 3. (a)** TM$_{10}$ mode resonant frequency, $\omega_S$, of conventional PPR given by (3), Josephson plasma frequency, $\omega_J$, given by (A5), and TM$_{10}$ mode resonant frequency, $\omega_{res}$, of TPPR given by (9) as functions of normalized dielectric spacer thickness $S/\xi_d$ around the Swihart-Josephson crossover. **(b)** TM$_{10}$ mode quality factor of conventional PPR, $Q_S$, given by (4), TM$_{10}$ mode ohmical quality factor of TPPR, $Q_\Omega$, from (10), quasiparticle tunneling quality factor of TPPR, $Q_{qp}$, from (10), and TM$_{10}$ mode total quality factor of TPPR $Q$ given by (10) as functions of $S/\xi_d$.